\documentclass[aps,pre,twocolumn,superscriptaddress,floatfix,showpacs]{revtex4}
\usepackage{graphicx}
\usepackage{bbm}
\usepackage[intlimits]{amsmath}
\usepackage{amssymb}
\newcommand{\be}{\begin{equation}}
\newcommand{\beq}{\begin{equation}}
\newcommand{\ee}{\end{equation}}
\newcommand{\bea}{\begin{eqnarray}}
\newcommand{\eea}{\end{eqnarray}}
\newcommand{\ba}{\begin{array}}
\newcommand{\ea}{\end{array}}

\begin{document}
\title{Fractal dynamics in chaotic quantum transport}

\author{V. Kotim\"aki}
\affiliation{Nanoscience Center, Department of Physics, University of
  Jyv\"askyl\"a, FI-40014 Jyv\"askyl\"a, Finland}
\author{E. R{\"a}s{\"a}nen}
\email[Electronic address:\;]{esa.rasanen@tut.fi}
\affiliation{Department of Physics, Tampere University of Technology, 
FI-33101 Tampere, Finland}
\affiliation{Nanoscience Center, Department of Physics, University of
  Jyv\"askyl\"a, FI-40014 Jyv\"askyl\"a, Finland}
\affiliation{Physics Department, Harvard University, 
Cambridge, Massachusetts 02138, USA}
\author{H. Hennig}
\affiliation{Physics Department, Harvard University, 
Cambridge, Massachusetts 02138, USA}
\author{E. J. Heller}
\affiliation{Physics Department, Harvard University, 
Cambridge, Massachusetts 02138, USA}

\date{\today}

\begin{abstract}
Despite several experiments on chaotic quantum transport in two-dimensional
systems such as semiconductor quantum dots, corresponding quantum 
simulations within a real-space model have been out of reach so far. 
Here we carry out quantum transport 
calculations in real space and real time for a two-dimensional 
stadium cavity that shows chaotic dynamics. By applying a large set 
of magnetic fields we obtain a complete picture of 
magnetoconductance that indicates fractal scaling. 
In the calculations of the fractality we use detrended fluctuation analysis
-- a widely used method in time series analysis -- and show its
usefulness in the interpretation of the conductance curves.
Comparison with a standard method to extract 
the fractal dimension leads to consistent results that, in 
turn, qualitatively agree with the previous experimental data.
\end{abstract}


\maketitle

\section{Introduction}

Since the pioneering works of Mandelbrot~\cite{Mandelbrot:1983uo}, 
fractal patterns have been found in a variety of 
objects in nature including, e.g., snowflakes, 
fern leaves, coastlines~\cite{Mandelbrot:1967uo,Meakin:1998te}, 
and even music~\cite{Voss:1975wm,Hsu:1990wv,Hennig:2011hq,Hennig:2012vl}. 
These self-similar (or self-affine) structures
were also found in many branches of chemistry and physics;
prominent examples are crystal growth and fractal surfaces, 
and transport in gold nanowires and electron 
``billiards''~\cite{Meakin:1998te,Michely:2004ua,Barnsley:2012wd,Sachrajda:1998wz,adam98,Hegger:1996vv,adam04,marlow,adam12}. 
In contrast with idealized mathematical fractals continuing to infinitely 
small scales, fractal scaling in nature has a lower and 
an upper limit.

While fractals found in nature are often well described 
by classical theories~\cite{Mandelbrot:1983uo,Meakin:1998te,Michely:2004ua,Barnsley:2012wd}, 
fractals have also been suggested to manifest in 
different quantum systems \cite{Berry:1996vd,Wojcik:2000uy,Hufnagel:2001vb,Benenti:2001to,Guarneri:2001ha,Louis:2000wg},
where a fundamental lower cutoff for fractal 
scaling is given by the Heisenberg uncertainty principle. 
In the case of transport through chaotic systems, 
such as chaotic electron billiards, both semiclassical~\cite{Ketzmerick:1996vj} 
(involving quantum interference) and classical 
mechanisms~\cite{Hennig:2007fs} for the emergence 
of fractal scaling have been proposed. 

For quantum systems with an underlying 
classically \emph{mixed} phase space with 
both regular and chaotic regions, 
a quantum graph model suggests a splitting
of the chaotic regime into two parts~\cite{Hufnagel:2001vb}: 
one part yields fractal conductance fluctuations while the 
other one leads to isolated resonances on small scales. 
These isolated resonances were later
shown to be associated with the eigenstates of a closed 
system~\cite{Backer:2002gk}. 
 
A stadium quantum billiard of charged particles is a generic 
chaotic system, whose underlying classical phase space is chaotic.
The phase space becomes mixed in presence of a 
(perpendicular) magnetic field. 
In the past two decades the system has been subject to 
several experiments~\cite{Marcus:1992ug,Micolich:2001vh,Takagaki:2000ud,Kuhl:2005fi,Sachrajda:1998wz}.
A typical setup consists of the two-dimensional (2D) electron gas 
(2DEG) in a semiconductor heterostructure, where
metallic gates are used to form the geometrical shape of
the ``billiards'' -- here called a quantum dot.
Alternatively, stadium billiards (and other chaotic systems) 
can be realized experimentally with microwave 
cavities \cite{Kuhl:2005fi}.

Dynamics in chaotic cavities has been extensively
studied with various theoretical methods including, e.g.,
random matrix theory ~\cite{Baranger:1994}, trajectory-based semiclassical theory ~\cite{Heusler:2006}, quantum mechanical kicked-rotor models~\cite{Casati:2000}, and tight-binding calculations~\cite{Louis:2000wg}. Semiclassical and random matrix theory have been used to investigate weak localization and Ehrenfest time effects~\cite{Jacquod:2006,Brouwer:2006}, while the kicked-rotor model and tight-binding studies have focused on the fractal structure of the quantum survival probability in chaotic cavities and the effect of changing the width of the output leads~\cite{Louis:2000wg}, respectively. Benenti {\em et al.} provide evidence for fractal fluctuations of the quantum survival probability in nonclassical situation of strong localization~\cite{Benenti:2001to}. However, to the best of our knowledge, none of the previous dynamical approaches have focused on conductance calculations in 2D chaotic cavities described by real-space grids in space and time.

It is worthwhile to notice that, in principle, the conductance problem of a chatic cavity can be treated within the conventional transport formalism, where the equilibrium current is obtained {\em time-independently}~\cite{transport}. In this approach, the coupling matrix of the cavity eigenstates and the lead states need to be evaluated. The most tedious part is an accurate and efficient treatment of the 2D eigenvalue problem for the chaotic cavity in real space and in the presence of the magnetic field. Recent progress has been made in this direction~\cite{perttu}, and such a conventional transport scheme is subject of future work. Nevertheless, as shown below, the present dynamical approach provides an efficient way to assess the conductivity and gives also additional information on time-dependent effects in the system.


In this work we calculate the fractal scaling of conductance 
fluctuations in an open {\em quantum} stadium billiard in
a full 2D model in real space and real time.
Our explicit solution of the time-dependent Schr\"odinger
equation for chaotic transport
goes beyond both the semiclassical treatment~\cite{Ketzmerick:1996vj}
and the above mentioned quantum graph model~\cite{Hufnagel:2001vb}.
We analyze the fractal scaling using two methods that originate 
from different fields of physics: the variation 
method~\cite{Meakin:1998te,Dubuc:1989vq,Sachrajda:1998wz} and
detrended fluctuation
analysis~\cite{Peng:1995wi,Kantelhardt:2001uk,PengDNA} (DFA). 
The variation method was used by Sachrajda {\em et al.}~\cite{Sachrajda:1998wz}
for the analysis of {\em experimental} magnetoconductance curves. We are
able to find a good agreement between theory and experiment,
both yielding a fractal dimension $D \sim 1.3$.


\section{Model and the computational scheme}

\begin{figure}
\includegraphics[width=1.0\columnwidth]{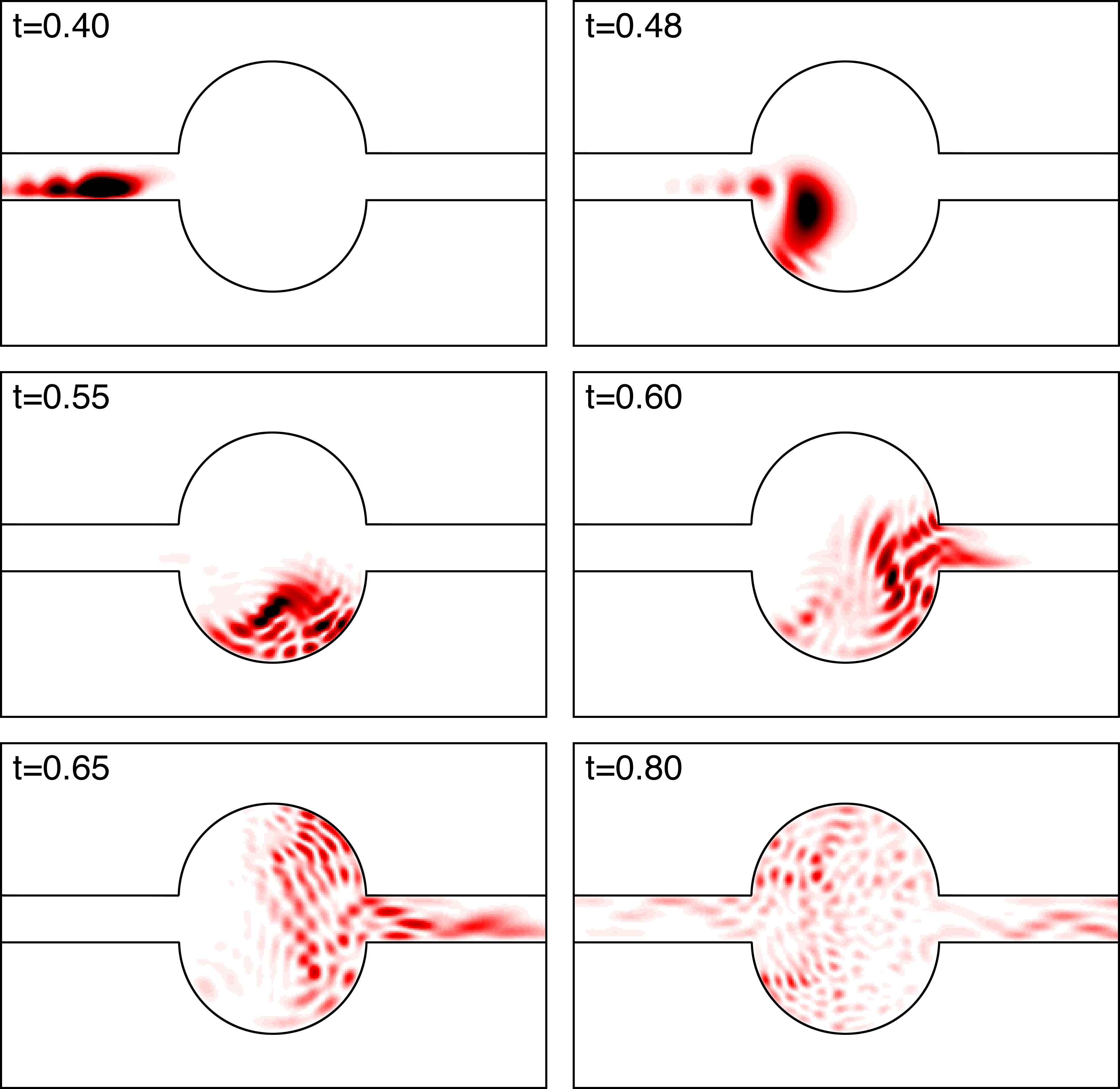}
\caption{Snapshots of the electron density in
the model stadium system (see text) during a transport 
simulation with the magnetic flux $\Phi/\Phi_0 = 20$. The input
and output leads extend further to the left and right.}
\label{fig1}
\end{figure}

We consider a model for semiconductor stadium device fabricated in 
the 2DEG of a AlGaAs/GaAs heterostructure similar 
to Ref.~\cite{Sachrajda:1998wz}.
The Hamiltonian describing our 2D system reads (in atomic units)
\begin{equation} \label{Ha}
\hat H = \frac{1}{2}\left[-i\nabla+\mathbf{A}(\mathbf{r})\right]^2+V_{\rm ext}(\mathbf{r},t),
\end{equation}
where the vector potential is given in the linear gauge, 
$\mathbf{A}(\mathbf{r})=(-By,0,0)$, to describe a
static and uniform magnetic field perpendicular to the plane.
During the time-propagation at $t>0$, the potential 
$V_{\rm ext}(\mathbf{r},t)$ consists of three parts: (i)
a stadium with radius $r=1$ and width $d=0.7$, (ii)
input and output leads of width $w=0.56$, and (iii)
a linear potential along the propagation direction in the first
two thirds of the input lead describing a source-drain voltage. The potential
has hard  boundaries with a depth $V_0=10000$ and the slope of the 
accelerating linear potential is $-100$. The central
part of the external potential is shown in Fig.~\ref{fig1}.
The input and output lead extend further to the left and right.

The {\em initial} state at $t=0$ is calculated by taking a small part 
of the input lead as a potential well. The resulting ground state of 
a single electron in the well is then used as an initial state for 
the time propagation. At $t>0$ the above described linear potential accelerates 
the wave packet across the system. For the time propagation 
we use a fourth-order Taylor expansion of the time-evolution operator. 
The {\tt octopus} code package~\cite{Octopus} 
is used in all the calculations.

We assess the conductance by calculating the integrated probability 
density in the output lead from
\begin{equation}
T(\Phi, t) = \int_{\rm output}\mathrm{d}\mathbf{r} | \psi (\Phi,\mathbf{r},t)|^2,
\end{equation}
where $\Phi$ is the fixed magnetic flux, given
above in units of the magnetic flux quantum $\Phi_0=h/e$. 
We call $T$ as a transmission factor assumed to be
proportional to the transmission coefficient available
in conventional transport theory. The validity of
the transmission factor in estimating
the {\em relative} conductivity as a function of an external 
parameter -- here the magnetic flux -- has been justified
in Ref.~\cite{rings}. Thus, we repeat the time-propagation
for different values of $\Phi$ to obtain the
magnetoconductance that can be compared with the
experiments in Ref.~\cite{Sachrajda:1998wz}.

It is important to note that our calculations allow
energy dispersion for the wave functions in the cavity, i.e.,
we describe nonstationary states. In this respect, our results
are not directly comparable to those of Ref.~\cite{Ketzmerick:1996vj}.
However, our transport approach qualitatively describes
an experimental situation to the extent that each value for
the magnetic field is treated equally, so that we can compare
the relative conductance as a function of $B$. Previously, a similar
approach has been used to assess
quantum conductance in quantum rings~\cite{rings} 
and Aharonov-Bohm interferometers~\cite{ab1}.

\section{Transport simulations}

In Fig.~\ref{fig1} we show snapshots of the electron 
density at different times at $\Phi = 20\Phi_0$ through the 
stadium. Approximately one half of the density is transferred
through and other half is either reflected back to the input
lead or confined in the stadium. As expected, the density
is scattered in the stadium in a chaotic fashion. The
size of the wiggles during the scattering depends on the
momentum -- the higher the momentum the higher eigenstates
are probed. We point out that the modes are not set
prior to the calculation, but the wave packet is freely
scattered and dispersed in the cavity. Here, we have chosen 
the initial momentum of the wave packet such that
considerable overlap is found with $\sim 50\ldots 100$
eigenstates of the stadium during the transport. 
This corresponds to considerable 
qualitative complexity in the propagated density, which,
as shown below, leads to a complex behavior of $T$.
On the other hand, the momentum is limited by the 
grid spacing of the simulation box -- all the nodes 
in the scattered wave packet should be accurately
described.

\begin{figure}
\includegraphics[width=1.0\columnwidth]{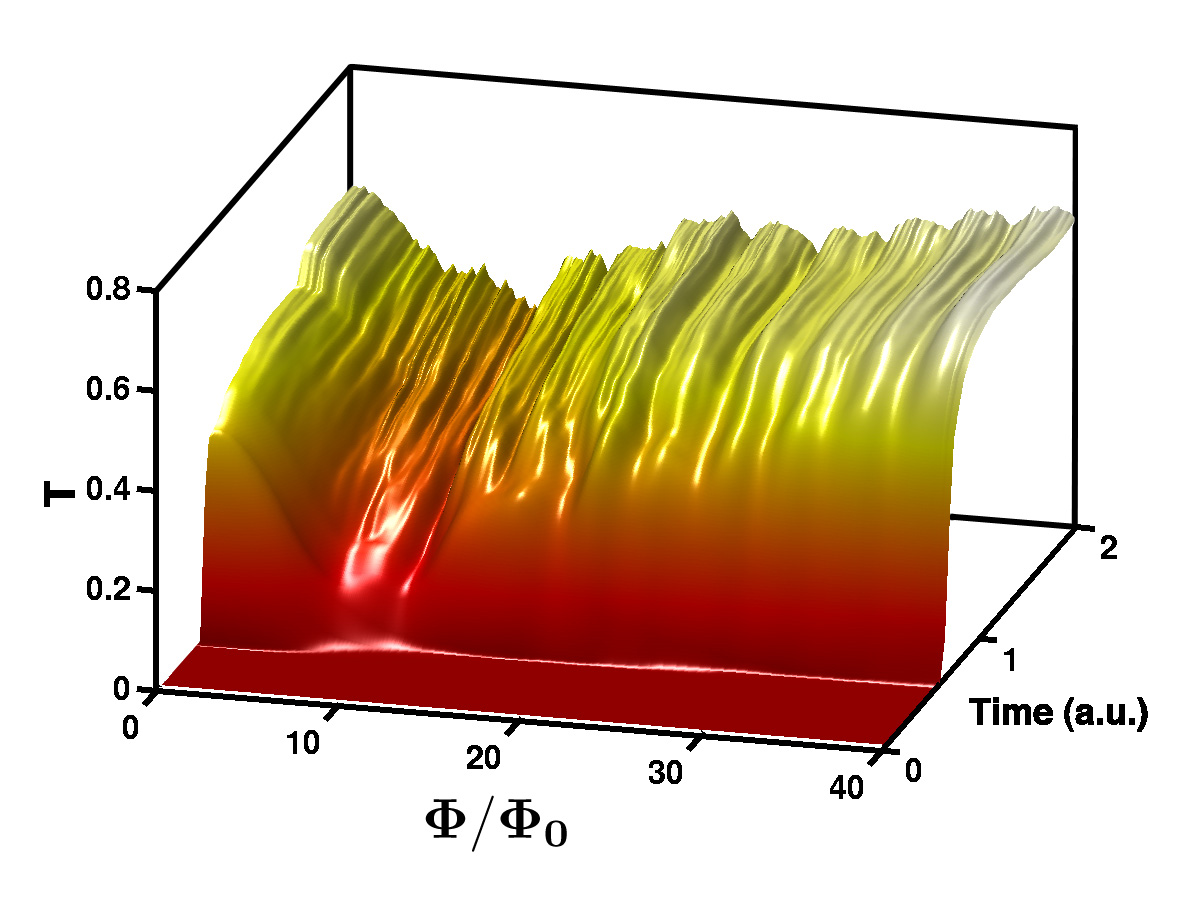}
\caption{Transmission factor as a function of time 
and magnetic flux through the stadium.}
\label{fig2}
\end{figure}

\begin{figure}
\includegraphics[width=1.0\columnwidth]{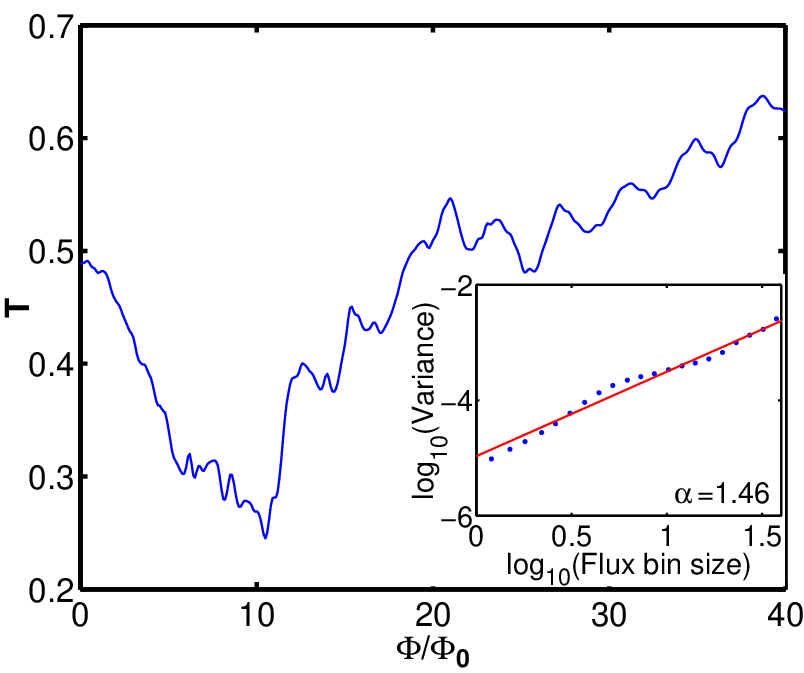}
\caption{Transmission factor as a function of 
the magnetic flux at $t=1.4$. The inset shows the 
scaling exponent $\alpha=1.46$ obtained from 
the DFA analysis.}
\label{fig3}
\end{figure}

A complete presentation of our transport results
is given in Fig.~\ref{fig2}, where the transmission factor 
$T$ is plotted as a function of both time and 
the magnetic flux. The figure consists of 401
respective time-propagations, each with a
fixed number of flux quanta $\Phi/\Phi_0$ ranging from 
zero to 40 in steps of 0.1. The flux range is qualitatively
similar to the experiment in Ref.~\cite{Sachrajda:1998wz}.
A complex magnetoconductance is formed if the propagation
time is larger than $\sim 1$. A cross section of the
conductance at $t=1.4$ is shown in Fig.~\ref{fig3}. 
We point out that due to the finite system size we are
not able to reach the {\em equilibrium} current and
thus find the absolute conductance. In practice, we 
stop the time-propagation immediately when the backscattering from
the walls of the calculation box becomes visible. Therefore, we 
consider fixed propagation times through the parameter
range of $\Phi/\Phi_0$. In other words, a fixed propagation
time is expected to treat all the values of $\Phi/\Phi_0$
equally in order to obtain the relative conductance $T$.

We first briefly consider the general trends in $T$ in Fig.~\ref{fig3}. 
As the flux is increased from zero the conductance decreases
mainly due to the vanishment of trajectories directly coupling
the left and the right lead. After reaching the minimum the
conductance generally becomes larger, which is due to
the increase of skipping orbits along the boundaries of the 
system. At large fields, interference effects play an important
role~\cite{vanhouten}. We point out, however, that the dynamics
is largely chaotic through the whole range of fluxes considered
here -- possibly only apart from the zero-flux limit.

Now, the essential question is whether the conductance
as a function of the magnetic flux shows fractal
characteristics. Moreover, it is interesting to 
consider how large propagation times are required
to find fractals.
This is analyzed in the following with two techniques:
the variation method~\cite{Meakin:1998te,Dubuc:1989vq,Sachrajda:1998wz}, 
and DFA~\cite{Peng:1995wi,Kantelhardt:2001uk,PengDNA}.

\section{Methods for fractal analysis}

\subsection{Variation method}

To extract the fractal dimension $D$ for a
mapping $f:\mathbb{R}\to\mathbb{R}$, the
domain of the given function is first divided into length 
intervals $\Delta x$.
The difference between the minimum and maximum of the 
function is calculated within every interval and added up. 
Note that the intervals are shifted window-wise accross 
the x-axis (not point-wise). 
In case of fractal scaling, the resulting sum is a power-law 
function of the interval length~\cite{Meakin:1998te,Dubuc:1989vq,Sachrajda:1998wz}:
\begin{equation}
\sum_i [ \max f(x) - \min f(x) ]_{|x-x_i|<\Delta x/2} \propto (\Delta x)^{-D+1}.
\end{equation}

\subsection{Detrended fluctuation analysis}

DFA is a standard method that was developed in the context 
of time-series analysis to study $1/f$ noise and 
long-range correlations~\cite{Peng:1995wi}
and has proven to be very reliable particularly in dealing 
with non-stationary time-series and trends in the 
data~\cite{Peng:1995wi,Kantelhardt:2001uk,Jennings:2004il,Ivanov:2009ei,Hennig:2011hq}.
It has also been used outside the time domain, e.g., to study the 
organization of DNA nucleotides~\cite{PengDNA}.
However, to our knowledge, DFA has not been applied to fractal 
conductance curves before, and the application of DFA to reproducible 
fractals is typically not straightforward.

The standard procedure of DFA consists of the following four
steps~\cite{Peng:1995wi,Kantelhardt:2001uk}: 
(1) integrating the time series, (2) dividing the series into windows 
of size $s$, (3) fitting with a polynomial $f_{s}(i)$ of degree $m=2\ldots 4$ 
that represents the trend in the window, and (4) 
calculating the variance with respect to the local trend $f_s(i)$
from
\begin{eqnarray}
F(s)&=&\left<(f(i)-f_{s}(i))^2\right> \nonumber \\
&=&\frac{1}{N-1}\sum_{i=1}^{N}\left(f(i)-f_{s}(i)\right)^2 \propto s^\alpha.\label{eq:dfa_fofs}
\end{eqnarray}

The key point in applying DFA to study conductance fluctuations 
is to relate the exponent $\alpha$ to the quantity of interest 
(here: the fractal dimension $D$). It is known that $D=2-\gamma/2$ 
with $\left<(\Delta G)^2\right> 
\propto (\Delta B)^\gamma$~\cite{Sachrajda:1998wz, Ketzmerick:1996vj}.
The latter is exactly step (4) of the DFA analysis above. We 
therefore omit step (1) and identify $\alpha=\gamma$, hence the 
fractal dimension reads $D=2-\alpha/2$.

\section{Results on the fractality}

In DFA, we apply quadratic detrending ($m=2$) to
our data in Fig.~\ref{fig3}. The inset shows the fitting of 
the data (solid line) at $t=1.4$ that yields
$\alpha=1.46$. This qualitatively agrees well with the
experimental result $\gamma=\alpha=1.38$ of
Sachrajda~\emph{et al.}~\cite{Sachrajda:1998wz}. 
The corresponding fractal dimension extracted from DFA 
is $D=1.27$. In comparison, the variation method yields
$D=1.32$ for our data, whereas the corresponding
experimental result -- obtained with the same method --
is $D=1.25$~\cite{Sachrajda:1998wz}. The expected error bars
for our results are discussed below. Nevertheless,
we find an excellent qualitative agreement of the results 
both regarding the different methods and comparison
with the experimental data. We point out that
our stadium model is similar to the experiment and the 
channel dimensions are also comparable.
According to our calculations, increasing the channel 
width from $0.56$ to $0.7$ leads to the same $D$ obtained 
in the variation method, whereas DFA yields a slightly
smaller $D$.

\begin{figure}
\includegraphics[width=1\columnwidth]{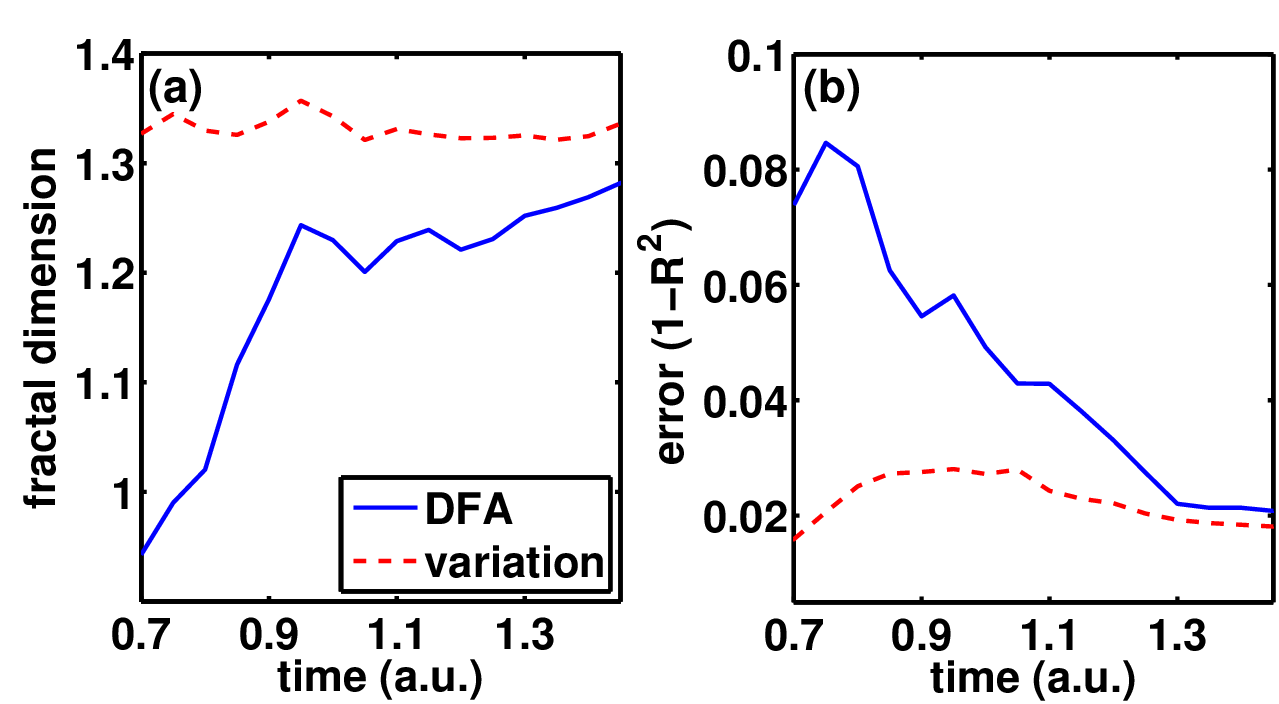}
\caption{(a) Fractal dimension $D$
calculated from DFA with the relation $D=2-\alpha/2$ 
and from the variation method, respectively, 
during the time-propagation. Note that the fractal
structure is developed only at $t\gtrsim 1$.
(b) Time-development of the error
in the fitting procedure at $t=0.7 \ldots 1.4$ (see text).}
\label{fig4}
\end{figure}

In Fig.~\ref{fig4}(a) we show the time-development
of the fractal dimension obtained from DFA and
the variation method, respectively. 
We point out that clear signatures 
of a fractal structure are developed only at $t\gtrsim 1$.
Nevertheless, $D$ converges during the time-propagation 
towards the values given above, and the quality of the
fitting in both methods improves as well. 
Figure~\ref{fig4}(b) shows the error in 
the fitting, ${\rm ERR}=1-R^2$, for
times $t=0.7\ldots 1.4$. Here
$R$ is the Pearson product-moment correlation 
coefficient of the log-log data. Thus, ERR measures
the linear fit quality such that ${\rm ERR}=0$
corresponds to exact linear behavior. The minimum
of the error is obtained at $t\approx 1.4$, which 
is the optimal time used above to determine $\alpha$ 
and $D$. 
At larger times with $t\gtrsim 1.4$ the error
increases due to back-scattering effects resulting
from the finite simulation box (see above).
In this way we are able to determine the range of 
validity in our scheme to calculate the fractal dimension. 

It is important to note that in addition to the numerical 
error of the fitting procedure (see above), the algorithms 
for fractal analysis have {\em internal} error bars 
analyzed in detail by Pilgram and Kaplan~\cite{Pilgram}.
For example, DFA results for the fractal scaling are expected 
to have a standard deviation of $\sim 15\%$ for data sets that 
are the of same size with ours. The results from the variational 
analysis are expected to contain similar deviations. 

Finally we point out that qualitatively similar fractal
dimensions have been obtained in various experiments
on billiard systems of different shapes~\cite{adam98,adam04,marlow}.
The dependence of $D$ on energy-level resolution
determined by experimental conditions has been discussed  
in several works~\cite{adamreview}. Moreover, the considerable
role of disorder in the modulation-doped 2DEG was recently
demonstrated~\cite{adam12}. However, in the same work it was
shown that {\em electrostatic doping} leads to reproducible properties
in thermal cycling. In view of these recent advances
it can be expected that ballistic transport properties 
of 2DEG billiard systems will be determined 
in forthcoming experiments with a high precision.
In turn, this motivates us to extend the applications of
the present method to various geometries.


\section{Summary}

In summary, we have calculated the time-evolution of a 
single-electron wave packet through a two-dimensional 
stadium-shaped cavity by solving the Schr\"odinger equation
in real time and real space. The relative conductance
has been calculated for a large set of magnetic fluxes
in order to analyze the fractal nature of the magnetoconductance.
We have found that the conductance shows clear indications for
fractal scaling. The fractal dimensions extracted from two
respective methods are consistent with each other.
Moreover, we have found an excellent qualitative agreement with previous 
experimental results. Our findings indicate that DFA suits well
for the analysis of fractal scaling in chaotic quantum transport.
Hence, we suggest to extend the use of the concept of data detrending 
(and hence DFA) to study fractal scaling of transport and other 
characteristics in chaotic (quantum) systems.

\begin{acknowledgments}
We thank Adam Micolich and Rainer Klages for very helpful comments
and discussions.
This work was supported by the Magnus Ehrnrooth Foundation, Wihuri Foundation, 
and the Academy of Finland. HH acknowledges funding through the 
German Research Foundation (DFG, grant no.~6312/1-2).
CSC Scientific Computing Ltd is acknowledged for 
computational resources. 
\end{acknowledgments}


\end{document}